\documentclass[10pt,twocolumn]{article}

\usepackage[utf8]{inputenc}
\usepackage[T1]{fontenc}
\usepackage[margin=0.75in, columnsep=0.25in]{geometry}
\usepackage{lmodern}
\usepackage{microtype}

\usepackage{authblk}

\usepackage{amsmath, amssymb, amsthm}

\theoremstyle{definition}

\usepackage{graphicx}
\usepackage{adjustbox}
\usepackage{xcolor}
\usepackage{caption}
\usepackage{subcaption}
\usepackage{float}
\usepackage{siunitx}
\usepackage{dirtytalk}
\usepackage{ragged2e}
\usepackage{fancyhdr}

\usepackage[numbers,sort&compress]{natbib}
\usepackage[colorlinks=true, linkcolor=blue, citecolor=blue, urlcolor=blue]{hyperref}

\title{\textbf{General formulation of an analytic, Lipschitz continuous control allocation for thrust-vector controlled rigid bodies}}

\author[1]{Frank Mukwege\thanks{\texttt{frank.mukwege@ulb.be}}}
\author[2]{Tam W. Nguyen\thanks{\texttt{nguyen.tamwilly.3e@kyoto-u.ac.jp}}}
\author[1]{Emanuele Garone\thanks{\texttt{emanuele.garone@ulb.be}}}

\affil[1]{Universit\'e Libre de Bruxelles, SAAS - Control and System Analysis Department, Avenue Franklin Roosevelt 50, CP 165/55, Brussels, Belgium}
\affil[2]{Kyoto University, Department of Electrical Engineering, 615-8510 Kyoto, Japan}

\date{}
\graphicspath{ {./images/} }

\fancypagestyle{plain}{%
  \fancyhf{}
  \fancyfoot[C]{Preprint submitted to Automatica}
  
}

\title{General formulation of an analytic and Lipschitz continuous control allocation for thrust-vectored controlled rigid-bodies\footnotemark[1]}

\begin{document}
\maketitle
\begin{abstract}                          
This paper presents a general framework for solving the control allocation problem (CAP) in thrust-vector controlled rigid bodies with an arbitrary number of thrusters. Two novel solutions are proposed: a closed-form, Lipschitz continuous mapping that ensures smooth actuator orientation references, and a convex optimization formulation capable of handling practical actuator constraints such as thrust saturation and angular rate limits. Both methods leverage the nullspace structure of the allocation mapping \textcolor{black}{to preserve control authority} while generating sub-optimal yet practical solutions. {The resulting approaches allow to arbitrary select a desired trade-off between Lipschitz continuity constants and energy efficiency loss}. The effectiveness and generality of the proposed framework are demonstrated through numerical examples on a marine vessel and an aerial quadcopter.
\end{abstract}

\section{Introduction}

Control allocation \textcolor{black}{(CA)} problems manifest when there exists an excess of control signals in comparison to the commanded forces and moments \cite{fossen_survey_2006}.  In aerospace and marine systems, control allocation techniques are employed to efficiently harness this redundancy of actuators in order to collectively generate the generalized forces $\tau_d$ commanded by a higher-level controller while fulfilling actuator constraints. 
In this work, a general framework for control allocation of overactuated rigid bodies equipped with vectored thrusters, i.e propulsion units capable of producing a thrust $T$ in a given direction is presented. Thrust vectoring is particularly useful to enable overactuation in systems while using less physical actuators on board. Indeed, one can mention nozzles used in military aircrafts and rockets, azimuth thrusters (electric or gasoline pods) of unmanned surface vessels \textcolor{black}{(USV)}  (Figure \ref{fig:ActuatorFrame}) or tiltable propellers of multi-rotor unmanned aerial vehicles \textcolor{black}{(UAVs)}. 

In the early 90s, most of the solutions for the control allocation problem (CAP) originated from the aeronautics community. These were mostly closed-form solutions based on the Moore-Penrose pseudo-inverse and the exploitation of the null-space (\cite{durham_constrained_1993,bordingnon_closed-form_1995,bordignon1996constrained,nullspace}). While those pseudo-inverse solutions are fast and easy to implement, they are sensible to the singularities that induce loss of control authority, i.e the ability to produce the required control actions in all directions of the wrench space $\mathcal{T}_d$ at all times. \textcolor{black}{In this work, we refer to and classify singularities into three categories: (i) \textit{rank-deficient singularities}, where specific actuator orientations cause the allocation mapping to lose full row rank, compromising control authority; (ii) points in the wrench space that induce discontinuous orientation references, leading to transient loss of control authority due to finite servo rate limits; and (iii) \textit{kinematic singularities}, actuator orientations where the chosen angular parametrization becomes mathematically undefined (e.g., gimbal lock). All those singularities induce a loss of the overactuation of the system and often can lead to instability of the closed-loop.} Although these singularities are problematic, they are also usually related to energy efficiency. Therefore, handling these singularities involves a trade-off between optimality in terms of energy consumption and stability of the closed-loop system. 
\subsection{Related work and state of the art}
 In the marine community, the CAP has been usually addressed by resorting to online optimization. As an example, the seminal work \cite{johansen2004constrained} presents a sequential quadratic programming strategy, solving a local approximation of a nonlinear CAP. To handle singularities, a penalty term is incorporated into the cost function, effective only against linearly dependent configurations and only possible when dealing with 3 or more actuators. In \cite{su_nullspace-based_2021}, a similar approach is presented for the control of overactuated UAV platforms by extending \cite{johansen2004constrained}'s local approximation of the CAP by exploiting the nullspace to reduce the linearization error. In \cite{ruth2007thrust}, a convex quadratic representation of the CAP is presented, accounting for thrust saturation and angular rate constraints. This solution is based off Sordalen's extended representation \cite{SORDALEN19971223}. A novel constrained control allocation method is detailed in \cite{scibilia_constrained_2012}, emphasizing a predictive methodology to compute the subsequent set of actuator inputs at each sample. The work by \cite{CartThrustAlgo} introduces a Cartesian thrust allocation approach, addressing CAP for \textcolor{black}{3 degrees of freedom (DOF) }marine vessels with a diverse range of actuators, including tunnel thrusters or variable direction thrusters. Although constraints such as turn rate limits and thrust saturation are considered, the singularity avoidance solution proposed is only applicable for systems with more than 2 actuators. In \cite{torben_control_2020}, an algorithm for control allocation in double-ended ferries with azimuth thrusters is presented. This method reformulates the problem as a scalar-bounded optimization problem, facilitating efficient online resolution. \textcolor{black}{In \cite{santos_analysis_2022} and \cite{santos_fast_2021}, a fast control allocation method is proposed by decomposing the nonlinear control allocation problem into a set of linear sub-problems. This approach ensures rapid convergence and low computational cost, as validated by simulations and embedded-platform experiments.} In predictive control allocation, following nullspace approaches from the aerospace community of the 90s, \cite{nguyen_nullspace-based_2023} propose an optimal predictive control scheme for the manipulation of a ground-anchored object by a quadcopter. In \cite{nguyen_kernel-based_2025}, the authors presents a predictive CA method for a wide class of thrust vectoring nonlinear systems. \textcolor{black}{Finally, the CAP is often solved by the high-level controller as in Model Predictive Control (MPC). As introduced in \cite{veksler_dynamic_2016}, nonlinear MPC (NMPC) is an efficient solution to solve the dynamic positioning (DP) problem for surface vessels. The optimal problem is written to consider at the same time the CAP and the DP problem. In \cite{martinsen2019autonomous}, autonomous docking of a vessel is performed using NMPC that handles actuator limits as constraints and singularities with the penalty of \cite{johansen2004constrained} within the cost function.} 
 
 \textcolor{black}{Instead of solving an optimization problem at each time instant, the CAP can also be solved with a dynamic approach with updating policies (\cite{tjonnas_adaptive_2008}). In \cite{casavola_fault-tolerant_2010}, the framework was extended to fault-tolerant allocation for over-actuated systems. More recently, in \cite{tohidi_adaptive_2020}, an adaptive CA for uncertain over-actuated systems equipped with saturating actuator was proposed. The method distinguishes itself by handling uncertainty without estimating it. 
 While optimization-based and adaptive methods are standardized approaches in CA, machine learning-based methods started emerging recently. In \cite{raghunathan_design_2023}, a multi-layered autoencoder architecture is proposed for solving a nonlinear control allocation problem.} In \cite{skulstad2023constrained}, a novel constrained control allocation method for ships with azimuth thrusters using deep neural networks in an autoencoder architecture is proposed for a dynamic positioning context. This approach has been also investigated within the aerospace community as \cite{huan2018constrained,chen2016constrained} applied this solution to respectively a near-space vehicle and a space re-entry vehicle. \textcolor{black}{A similar approach using deep autoencoders was proposed in \cite{lan_deep-auto-encoder_2025} for a small over-actuated spacecraft. The interest of the method lies within its optimal energy and actuators effort distribution}.  
 
 \textcolor{black}{A suitable solution to the CAP has therefore to fulfill the controller demands while avoiding singularities and enforcing the actuator constraints.}
\textcolor{black}{Thus, this paper proposes two control allocation strategies tailored to thrust-vector controlled vehicles. While the results in \cite{MukwegeGaroneI} were derived for a surface vessel equipped with two azimuth thrusters in a specific configuration, the present work generalizes the framework to generic rigid-bodies with an arbitrary number of vectored thrusters placed at arbitrary locations. The contributions are summarized as follows:  
\begin{itemize}
    \item The first solution is a Lipschitz continuous closed-form solution to the unconstrained CAP \textcolor{black}{avoiding singularities} and offering robust stability guarantees;  
    \item The second solution is optimization-based, reshaping the constrained CAP into a convex form, and capable of effectively  \textcolor{black}{managing singularities} along with common actuator constraints such as thrust rate, saturation and angular rate.
\end{itemize}}  
\textcolor{black}{Both solutions introduces a tunable trade-off between the rate of change of the actuator references and the power consumption}. These 2 results will be illustrated by numerical examples on a surface vessel with azimuth thrusters and on an aerial quadcopter with tilting rotors.
\section{Problem Description}\label{sec: problem statement}

Let a rigid-body be characterized by $n$ degrees of freedom (DOF) and $m$ vectored thrusters and let a body frame \textbf{ $\boldsymbol{B}$ } be associated to it.  \\

\begin{figure}[h!]
    \centering
    \includegraphics[width = 0.5\columnwidth]{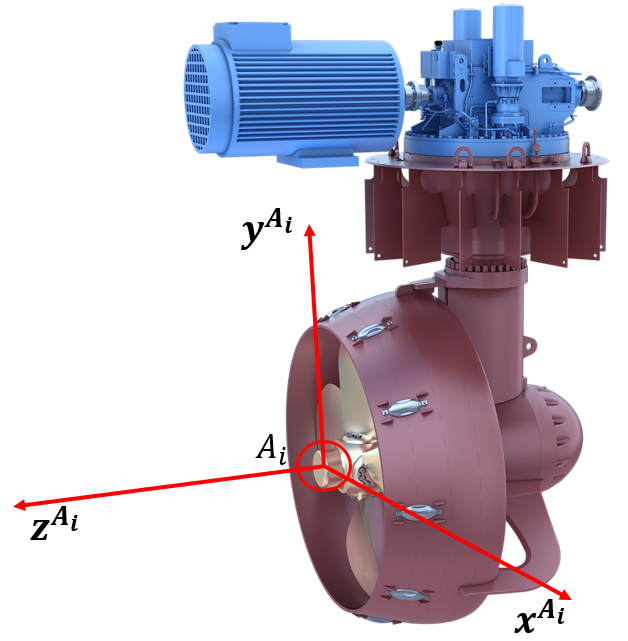}
    \caption{Example of an actuator frame for an azimuth thruster from \cite{kongsberg}}
    \label{fig:ActuatorFrame}
\end{figure}

For each $i^{th}$ actuator, consider a reference frame \textbf{ $\boldsymbol{A_{i}}$}, centered in the actuator and its $z$-axis points in the direction of the produced thrust $T$ (See Figure \ref{fig:ActuatorFrame}).
The forces and torques produced by this actuator in  the frame $\boldsymbol{A_{i}}$ are \begin{equation}
\boldsymbol{f}_{\boldsymbol{A_{i}}}=\left[\begin{array}{c}
0 \\
0 \\
1\\
\end{array}\right] T_i, \quad \boldsymbol{\tau}_{\boldsymbol{A_{i}}}=\left[\begin{array}{c}
0 \\
0 \\
-s_i \kappa_{d_i}
\end{array}\right] T_i,
\end{equation}\label{eq:Bryan} where $T_i \in \mathbb{R}^+$ is the thrust produced by the $i^{th}$ actuator. In the case of propellers  $\kappa_{d} \in \mathbb{R}^{+}_0$ is a proportional factor between the thrust and reaction torque depending on the application (e.g for helicopters, $\kappa_{d_i}$ is the length of the propeller blade), $s_i$ = ${\pm 1}$ represents the spin direction which implies the orientation of the reaction torque produced by the rotation of the propeller. For other type of thrusters (e.g. nozzles) which do not produce reaction torque, $\kappa_{d_i},s_{i}=0$. Note that in this work, unidirectional thrusters are assumed and therefore $T_i\geq 0$.

The forces and torques produced by the $i^{th}$ actuator in the body frame $\boldsymbol{B}$ are
\begin{equation}\label{eq:tau_i}
\tau_{i} = \left[\begin{array}{cc}
{R}_{\boldsymbol{A_i}}^{\boldsymbol{B}} & {0}_{3 \times 3} \\
{S}\left(p_{\boldsymbol{A_i}}^{\boldsymbol{B}}\right) {R}_{\boldsymbol{A_i}}^{\boldsymbol{B}} & {R}_{\boldsymbol{A_i}}^{\boldsymbol{B}}
\end{array}\right] \left[\begin{array}{c} 0\\ 0\\ 1\\ 0\\ 0\\-s_i \kappa_{d_i}\end{array}\right] {T_i},
\end{equation} 
where $R_{\boldsymbol{A_i}}^{\boldsymbol{B}}$ is the $\mathrm{SO(3)}$ rotation matrix from $\boldsymbol{A_{i}}$ to $\boldsymbol{B}$, where $p_{\boldsymbol{A_i}}^{\boldsymbol{B}}$ is the position of $\boldsymbol{A_i}$ in the reference frame $\boldsymbol{B}$ and   
${S}\left({p}_{\boldsymbol{A_i}}^{\boldsymbol{B}}\right)$ \textcolor{black}{the skew-symmetric matrix that represents the cross product operator of any vector with $p_{\boldsymbol{A_i}}^{\boldsymbol{B}}$}. With $p_{\boldsymbol{A_i}}^{\boldsymbol{B}}$ = $[p_x,p_y,p_z]$ : 
\begin{equation}
{S}\left({p}^{\boldsymbol{B}}_{\boldsymbol{A_i}}\right)=\left[\begin{array}{rcc}
0 & -p_z & p_y \\
p_z & 0 & -p_x \\
-p_y & p_x & 0\\
\end{array}\right].
\end{equation}
Through some manipulation of \eqref{eq:tau_i}, it is possible to rewrite $\tau_{i}$ in the compact form
\begin{equation}\label{eq:tau_i2}
\tau_{i} =M_i  \hat{v}_i T_i
\end{equation}\label{eq:CompactForm}
where $M_i \in \mathbb{R}^{6\times3}$ is defined as \begin{equation}\label{eq:MIMatrix}
    M_i =  \left[\begin{array}{c}
I_{3 \times 3} \\
{S}\left({p}^{\boldsymbol{B}}_{\boldsymbol{A_i}}\right) - s_i \kappa_{d_i} I_{3 \times 3}
\end{array}\right],
\end{equation}$\hat{v}_i \in \mathbb{S}^2$ is a $3 \times 1$ unit vector defined as the third column of the rotation matrix $R_{\boldsymbol{A_i}}^{\boldsymbol{B}}$ and represents the direction towards which the actuator is pointed. At this point, by summing all $m$ $\tau_{i}$, it is possible to get the effect of all the actuators on the rigid body as
\begin{align}
\tau =&\sum_{i=1}^m M_i  \hat{v}_i T_i \label{Total} \\ 
&= M \text{diag}_{3}\left(\left[ \begin{array}{c} \hat{v}_1 \\ \dots \\ \hat{v}_m  \end{array}\right]\right)\left[ \begin{array}{c}  T_1 \\ \dots \\  T_m \end{array}\right], \label{eq:3DGenForces}
\end{align}
where $\tau =  [F_{x},F_{y},F_{z},\tau_{x},\tau_{y},\tau_{z}]^T$ \textcolor{black}{is called the $6 \times 1$ \textit{generalized forces} vector}, the ${6\times 3m}$ static allocation matrix $M$ is 
\[
M=\resizebox{0.9\columnwidth}{!}{$\left[ \begin{array}{ccc}
I_{3 \times 3} & \dots & I_{3 \times 3}\\
{S}\left(p_{\boldsymbol{A_1}}^{\boldsymbol{B}}\right) - s_1 \kappa_{d_1} I_{3 \times 3} & \dots &  {S}\left(p_{\boldsymbol{A_m}}^{\boldsymbol{B}}\right) - s_m \kappa_{d_m} I_{3 \times 3}
\end{array}\right]$}.
\]
We define also $\text{diag}_{3} : \mathbb{R}^{3m\times1} \rightarrow \mathbb{R}^{3m\times m}$ is a matrix operator as follows 
\[
\resizebox{0.8\columnwidth}{!}{$\displaystyle\text{diag}_3\left(\left[ \begin{array}{c} \hat{v}_1 \\ \dots \\ \hat{v}_m  \end{array}\right]\right)=\left[\! \begin{array}{cccc} \hat{v}_1 & 0_{3\times1} & ... & 0_{3\times1}\\
0_{3\times1} & \hat{v}_2& ... & 0_{3\times1} \\
0_{3\times1} & 0_{3\times1} & ... & \hat{v}_m  \end{array}\!\right],$}
\] \textcolor{black}{where $\hat{v}_i$ is as introduced, the orientation of the $i^{th}$ thruster, and $0_{3\times1}$ is the $3\times1$ zero column vector}.

With thrust vectoring, it is possible to vary not only the thrust $T_i$ but also the orientation of the actuator $\hat{v}_i \in \mathbb{S}^2.$ \textcolor{black}{The orientation of the actuators is  governed by servos that can move around two degrees of freedom: the azimuth $\beta_i \in [-\pi,\pi)$, as the angle in the $x^{\boldsymbol{B}}y^{\boldsymbol{B}}$ plane, and the elevation $\alpha_i\in [0,\pi)$, as the angle in the $y^{\boldsymbol{B}}z^{\boldsymbol{B}}$ or $x^{\boldsymbol{B}}z^{\boldsymbol{B}}$ depending on the azimuth.} Accordingly each vector  $\hat{v}_i$ can be parameterized as 
 \begin{equation} 
 \label{eq:OrientationVector}
\hat{v}_i(\alpha_i,\beta_i)= \left[ \begin{array}{c} sin(\alpha_i)cos(\beta_i) \\ sin(\alpha_i)sin(\beta_i)  \\ cos(\alpha_i)  \end{array}\right].
 \end{equation}
At this point it is possible to rewrite \eqref{eq:3DGenForces} as
\begin{align}
\tau &= M  \mathrm{diag}_3 \left( 
\begin{bmatrix}
\hat{v}_1(\alpha_1,\beta_1) \\
\vdots \\
\hat{v}_m(\alpha_m,\beta_m)
\end{bmatrix}
\right)
\begin{bmatrix}
T_1 \\
\vdots \\
T_m
\end{bmatrix} \label{eq:Mappa} \\
&= J(\theta) T, \label{eq:Jtheta}
\end{align}

where $\theta=[\alpha_1, \beta_1, \dots, \alpha_m, \beta_m]^T,$ $T=[T_1,\dots,T_m]^T$ and the allocation matrix  $J(\theta)$ is defined as 
\begin{equation} 
J(\theta) = M diag_3\left(\left[ \begin{array}{c} \hat{v}_1(\alpha_1,\beta_1) \\ \dots \\ \hat{v}_m(\alpha_m,\beta_m)  \end{array}\right]\right).
\end{equation}\label{eq:MappingNL}
In this setting, the  Control Allocation Problem (CAP) can be defined as follows:

\textbf{Control Allocation Problem (CAP) -} Given a desired \textcolor{black}{generalized forces} vector $\tau_d$, find actuators orientations $\bar{\theta}$ and thrusts $\bar{T}$ such that 
$\tau_d = J(\bar{\theta}) \bar{T}.$

Many  variants of these problems can be formulated depending on possible different cost functions and constraints on the input as presented in \cite{johansen2004constrained}.
One of the basic variants is to choose, among the possible solutions, the one that minimizes some norm of the thrust while $T$ and $\theta$ are possibly subject to constraints. The basic case is the case where one wants to minimize a quadratic cost of the thrust subject to positivity constraints of the thrust. In this case it is convenient to rewrite \eqref{eq:Jtheta} in the so-called \textit{extended representation} (\cite{SORDALEN19971223}) 
\begin{equation}\label{eq:extendedMap}
\tau =   M F,  
\end{equation}
where $F=[F_1 F_2 ... F_m]^T$ and 
$
F_i = \hat{v}_i(\alpha_i,\beta_i) T_i
$. Doing so, the solution that minimizes the 2-norm of $F^*$ is obtained using the Moore-Penrose pseudo-inverse of $M$ 
\begin{equation}\label{eq:pseudoInv}
    F^* = M^{\dagger}\tau_d.
\end{equation}
Once the vector $F^*$ is obtained, the optimal thrust $T_i^*$ and orientations of each thrusters $\alpha_i^*, \beta_i^*$ are
\begin{align}
T_i^* &= \left\| F_i^* \right\|, \\
\beta_i^* &= \mathrm{atan2}(F_{y_i}, F_{x_i}),\label{eq:azimuth} \\
\alpha_i^* &= \mathrm{atan2} \left(\sqrt{F_{y_i}^2 + F_{x_i}^2} , \,F_{z_i}\right).
\end{align}
\textbf{Remark 1 -} Note that, even in the case F is subject to linear equality constraints in the form $A F = b$, the optimal solution of this problem can be computed in closed form as
\begin{equation}\label{eq:pseudoInv2}
        F^* = \left[ \begin{array}{c} M \\ A\end{array}\right]^{\dagger}\left[ \begin{array}{c}\tau_d \\ b \end{array}\right].
\end{equation}
This result is particularly useful in the case some actuators can be oriented only in one degree of freedom. For instance if only the elevation $\alpha_i$ can be modified and the azimuth is $\beta_i=0,$ this translates into the constraint
$
[0\,\,\, 1\,\,\,0]\, F_i = 0. \hfill \Box$
\\

\textbf{Remark 2 -} In some cases one might be interested in controlling the generation of forces and torques only along certain direction and the effect in others is not of interest. To tackle this case, a matrix $P \in \mathbb{R}^{6 \times l}$ can be used as a basis for the subspace in which the torques/forces needs to be generated.
For instance in the case of a vessel for which the forces and torques of interest are only on the x-y plane, $l = 3$ and $ P$ is
\begin{equation}\label{eq:TructatingP}
P=\left[ 
\begin{array}{ccc}
e_1 & e_2 & e_6 \\
\end{array}
\right],
\end{equation}
where $e_i \in \mathbb{R}^6$ is the $i^{th}$ vector of the canonical basis.
The matrix $P$ can be then used to ``cancel'' the unnecessary lines of \eqref{eq:extendedMap} obtaining $P^T \tau =  P^T M F.$ Accordingly, in this case, the optimal solution of the CAP is
\begin{equation}\label{extendedMap3}
F^*=(P^T M)^{\dagger}P^T \tau.
\end{equation}
$\hfill \Box$ 

The main drawback of the pseudo-inverse approach is that, although very easy to compute and to implement, it presents discontinuities in the orientations $\theta$. Since these angles are governed by servos that cannot instantaneously switch from one position to another, this is a major problem 
which can potentially lead  to undesired behaviors such as reduced bandwidth of the actuation, consequent undesired oscillations or even instability.


A conventional approach in the literature to address this problem is to rewrite the CAP as a nonconvex optimization problem to be solved at discrete time intervals (see e.g. \cite{johansen2004constrained,torben_control_2020}).
In our early paper (\cite{MukwegeGaroneI}) we proposed a solution to this angular discontinuity problems for the specific case of a surface vessel with two azimuth thrusters. Such solution drastically departs from the existing literature as it provides  a systematic methodology to build sub-optimal closed-form mappings that are Lipschitz continuous. We also show that, with some manipulation, the same results can be used to better handle all types of singularities and that moreover this can be done using convex optimization.


\section{Lipschitz continuous solution of the CAP in closed-form}\label{sec:Lips}
All the solutions $F \in \mathbb{R}^{3m\times1}$ to equation \eqref{eq:Jtheta} can be written as the optimal solution $F^*$ plus a vector in the null-space of the matrix $M,$ i.e.
\begin{equation}\label{AllSolutions} 
    F = M^{\dagger}\tau_d + z,
\end{equation} 
where $z\in Ker\{M\}$ and $M^{\dagger} \in \mathbb{R}^{3m\times6}$.
The main idea of the proposed solution is to use the vectors of the nullspace to ``smooth" the optimal mapping \eqref{eq:pseudoInv}. In particular the idea is to build a suboptimal mapping in the form
\begin{equation}\label{eq:OurSolutionLCAP}
    F = F^{*} + s(\tau_d),
\end{equation} 
where $F^{*} = M^{\dagger}\tau_d$ and $s(\tau_d)=[s_1^T(\tau_d),\dots,s_m^T(\tau_d)]^T$ is a function 
$ s(\cdot): \mathbb{R}^{6} \rightarrow Ker\{M\}$ that generates a smoothing vector designed to ensure the continuity of the vector $\theta$. This condition translates the continuity of $\hat{F}$, vector of the orientation of all the actuators defined as $ \hat{F} =  [\hat{F}_1,\dots,\hat{F}_m]$  with $ \hat{F}_j\in \mathbb{R}^{3\times1}$  :
\begin{equation}\label{eq:unitVector}
\hat{F}_j = \frac{F_j}{\|F_j\|}, j=1,...,m.
\end{equation}
The following Proposition gives sufficient conditions on $s(\tau_d)$ to build such a smoothing
\\
\\
\textbf{Proposition 1 -} Let us consider $\nabla$ the Clarke generalized gradient operator as defined in \cite{Clarke1990}, any compact domain $\mathcal{T}_d \subset \mathbb{R}^{6}$ and the function $s(\tau_d)\in Ker\{M\}, \forall \tau_d \in \mathcal{T}_d$ be such that:
\begin{itemize}
    \item \textbf{C1.} Its gradient is bounded for any bounded $\tau_d$, i.e., there exists \textcolor{black}{$\varepsilon_1 > 0$ with $\varepsilon_1 = \max_{\tau_d \in \mathcal{T}_d} \|\nabla s_{j}(\tau_d)\|$} such that:
    \[
    \|\nabla s_{j}(\tau_d)\|  \leq \varepsilon_1, \quad \forall j=1,\dots,m.
    \]
    
    \item \textbf{C2.} There exists a scalar $\varepsilon_2 > 0$ such that for any $\tau_{d} \in \mathbb{R}^{6}$:
    \[
    \| F^*_j + s_j(\tau_d) \| \geq \varepsilon_2, \quad \forall j=1,\dots,m.
    \]
\end{itemize}
then the mapping \eqref{eq:OurSolutionLCAP} is a solution of the CAP that ensures Lipschitz continuity of $\hat{F}_j, j=1,...,m$ on any compact domain $\mathcal{T}_d$. Moreover, for any compact $\mathcal{T}_d$, an upper-bound of the Lipschitz constants for each actuator orientation is 
\begin{equation}\label{Lips}
L^j \leq \frac{2}{\varepsilon_2} \left( \|M_j^\dagger\| + \varepsilon_1 \right), j = 1, \dots, m,
\end{equation}
where $M_j^\dagger \in \mathbb{R}^{3\times6}$ is the sub-matrix of $M^\dagger = [M_1^\dagger, \dots, M_m^\dagger]^{T}$ associated with the $j^{th}$ actuator.\\
\textbf{ \textit{Proof} -}  To prove the statement, it is enough to show that the jacobian of $\hat{F}_j$ is bounded. Using the formula of derivative of a normalized function from \cite{kraemer2007kernelizing} and \cite{kraemer2011degrees}, the jacobian of $\hat{F}_j$ is
\begin{align}
\nabla\hat{F}_j &= \nabla \left( \frac{F_j}{\|F_j\|} \right) \\
&= \frac{1}{\|F_j\|} \left( I_{3 \times 3} - \frac{F_j F_j^T}{F_j^T F_j} \right) \nabla F_j. \label{eq:RHSProof}
\end{align}
The norm of \eqref{eq:RHSProof} can be upper-bounded as 
$\hat{F}_j$ 
\begin{align}\label{eq:RHSProofNorm}
\|\nabla\hat{F_j}\| 
&\leq  \frac{1}{\varepsilon_2}\left(  1 + \left\lVert \frac{F_j F_j^T}{F_j^T F_j} \right\rVert \right)  \| \nabla F_j \|. 
\end{align}
The proof is completed by noticing that for any $\tau_d \in \mathcal{T}_d:$
\begin{itemize}
\item $\left\lVert \frac{F_j F_j^T}{F_j^T F_j} \right\rVert =1;$ 
\item $\|\nabla(F_j^* + s_j(\tau_d))\| = \|M_{j}^{\dagger} + \nabla s_{j}\| \leq \|M_{j}^{\dagger}\| +\varepsilon_1 $;
\end{itemize}
which finally implies : 
\begin{equation}\label{eq:upper-bound}
\|\nabla\hat{F_j}\| \leq \frac{2}{\varepsilon_2}
    (\|M_{j}^{\dagger}\| +\varepsilon_1).
\end{equation}
$\hfill \Box$
\\
Several ways to build smoothing mapping $s(\tau_d)$ exist. The following proposition gives a systematic way to build a simple smoothing function \\
%
\textbf{Proposition 2 - }Let a vector $K_{b}=[K_{b_1}^T,...,K_{b_m}^T]^T \in Ker\{M\}$ exists such that $\|K_{b_i}\|\geq 1, i=1,...,m.$ Let also $K_{i} \in \mathbb{R}^{3 \times 3}$ be an orthonormal basis such that the first vector of the basis is $K_{b_i}/\|K_{b_i}\|.$ Then the smoothing mapping 
\begin{equation}
    s(\tau_d) = K_{b} b(\tau_d)
\end{equation}
satisfies the conditions of Proposition 1 for any scalar positive function $b(\cdot):\mathbb{R}^{6} \rightarrow \mathbb{R}^+_{0} $ which satisfies the following two equations
\begin{equation}\label{eq:btauGradient}
\hspace{-0.5 cm} \| \nabla b(\tau_d) \| \leq \frac{\varepsilon_1}{\|K_{b}\|}\, ,
\end{equation}
\begin{equation}\label{eq:btauMechanism}
\begin{array}{l}
\hspace{-0.5 cm} b(\tau_d) \geq \max_{i=1,...,m}\frac{\varepsilon_2 - e_1^T K_{i}^{-1} F_i^*}{\|K_{b_i}\|} \\ 
\hspace{1 cm} \text{if} \min_{i=1,...,m} \left\lVert \left[\begin{array}{c} e_2^T \\ e_3^T \end{array} \right] K_{i}^{-1}F_i^*\right\rVert < \varepsilon_2,
\end{array}
\end{equation}
where $e_i$ is the $i^{th}$ vector of the canonical basis. \\
\textbf{ \textit{Proof} -} Condition \textbf{C1 }from Proposition 1 simply follows by substitution. 
Condition \textbf{C2} from Proposition 1 is fulfilled by equation \eqref{eq:btauMechanism}. Indeed, a smoothing action will appear along the nullspace direction $K_{b_i}$ of the $i^{th}$ actuator that produces orthogonally to this direction, a force that falls below $\varepsilon_2$. This is done by analysing the result $ \left\lVert \left[\begin{array}{c} e_2^T \\ e_3^T \end{array} \right] K_{i}^{-1}F_i^*\right\rVert < \varepsilon_2$. This enables to counteract all sources of discontinuities in the actuators orientation as the action along the nullspace ensures a systematic profile of $\hat{F}$. Away from the singularities, the solution $F$ is mainly based of $F^*$ and provides continuous orientation by definition. Close to the singularities, the null-space action of $K_bs(\tau_d)$ pushes away from them by modifying the projected forces vector $F$. All this ensures that some thrust gets always produced and this fulfills Condition \textbf{C2} : $\| F^*_j + s_j(\tau_d)\| \geq \varepsilon_2; \; \forall j=1,...,m$. With this result, a finite value of the Lipschitz constant \eqref{Lips} of the actuator orientation is ensured as the norm of $\|F\|$ will never be zero.  Therefore, any function $b(\tau_d)$ defined as \eqref{eq:btauGradient}-\eqref{eq:btauMechanism} fulfills the conditions of Proposition 1.
 $\Box$ 
 
\textbf{Remark 3 - } Although not needed for Proposition 2, in practice it is convenient to add a third condition to $b(\tau_d)$ that is 
\begin{equation}
\begin{array}{l}
\hspace{-0.5 cm} b(\tau_d) = 0, \,\,\, \text{if} \, \min_{i=1,...,m} \left\lVert \left[\begin{array}{c} e_2^T \\ e_3^T \end{array} \right] K_{i}^{-1}F_i^*\right\rVert >> \varepsilon_2.
\end{array}
\end{equation}
which ensures that under the condition that all actuators have components orthogonal to $K_b$ that are sufficiently large, the mapping \eqref{eq:OurSolutionLCAP} coincide with $F^*$  
$\hfill \Box$

\textbf{Remark 4 - } The vector $K_b$ can be interpreted as a ``rest configurations" of the actuators which produces zero total force/torque but where all the actuators are producing some thrust. 
Following the observation of Remark 3, if multiple choices are possible, it is convenient to select $K_{b}$ in such a way that is orthogonal to the ``typical" optimal configurations of the actuators. A possible way to compute $K_b$ is by solving the following optimization problem. 
\begin{equation}
\begin{aligned}\label{eq:BestDirection}
\min_{K_{b}} \quad & \|K_{b}\|^2_2 \\
\textrm{s.t.} \quad & K_{b} \in Ker(M)\\
  & \|K_{b_i}\| \geq 1, i=1,...,m\\
  &  K_{b_i}^T F_i^*(\tau^j) = 0, j=1,...,n_{typical}
\end{aligned}
\end{equation}
 where ${\tau}^j, j=1,..,n_{typical}$ are the ``typical" directions of large forces requested by the vehicle (e.g. forward force in $x$ for a ship, vertical thrust in $z$ for a multirotor, etc). \textcolor{black}{This additional constraint allows the designer to separate the ``useful" thrust that contributes to $\tau_d$ and ``smoothing" thrust that ensures Lipschitz continuity of the actuators orientation references.} Note that depending the number of actuators and the number of ``typical" configurations selected, the problem may become infeasible as it reduces the search space, thus the number of these ``typical" configurations must be chosen appropriately.  Although \eqref{eq:BestDirection} is not convex, the problem is still reasonably solvable offline. \textcolor{black}{Note that in the case of failure or disturbances acting on the system, depending how many extra actuators and smoothing direction are available, it is possible to pre-compute offline auxiliary directions to fall back depending of the scenario}.$\hfill \Box$
 
\textbf{Remark  5 - } An example of function $b(\tau_d)$ which satisfies Proposition 2 is 
\begin{equation}
    b\left(\tau_d\right)=max\left\{0, \; F_{ker_{i}}\right\} \text{g}\left(\tau_d\right) \; \forall i = 1 \dots m, \label{eq:Btaud}
\end{equation}

where $F_{ker_{i}}$, the force in kernel direction for the $i^{th}$ actuator, is defined as \begin{equation}
    F_{ker_{i}} = \frac{\varepsilon^*_2 - e_1^T K_{i}^{-1} F_i^*}{\|K_{b_{i}}\|} \forall i = 1 \dots m, \label{eq:Fker}
\end{equation} \\
 and $\text{g}(\tau_d) : \mathbb{R}^{6} \rightarrow \mathbb{R}^{+}_0$ is a sigmoid-like function such as :
\begin{equation}\label{eq:SmoothingFunction}
\resizebox{\columnwidth}{!}{$\displaystyle
\text{g}(\tau_d) = -k_{a} (\frac{2}{\pi} \operatorname{atan}\left(k_{b} (\min_{i=1,...,m}( \left\lVert \left[\begin{array}{c} e_2^T \\ e_3^T \end{array} \right] K_{i}^{-1}F_i^*\right\rVert)-\varepsilon^*_{2})\right)+1),$}
\end{equation}
with $k_{a}, k_{b} \in \mathbb{R}^{+}_0$ as tuning parameters allowing to choose the $\varepsilon_1$ value. $\varepsilon^*_2$ the is a smoothing threshold such that $\varepsilon_2 = \max_{\tau_d \in \mathcal{T}_d}\text{g}(\tau_d)\varepsilon^*_2$. In these notations, $\operatorname{atan}$ is the arctangent function.

\section{Dynamic optimization for constrained control allocation}
In many real applications, additional constraints to the positivity of the thrust might be imposed such as, e.g. thrust rate and saturation, limited servo velocities and pointing constraints. In these cases, analytical solutions cannot be easily computed and online optimization must be used \cite{johansen2004constrained,torben_control_2020,johansen_efficient_2005}. The typical approach presented in the literature to solve this problem while avoiding singularities in the mapping is to consider the problem in discrete time and to solve at each sampling time an optimization problem of the form:
\begin{equation}\label{eq:ClassicOptimization_cost}
    \min_{\theta,T,s} \, T^{T}WT + s^T Q s 
    + \frac{\varrho}{\epsilon +\det(J(\theta)J^T(\theta))}\\
    \end{equation}
\begin{equation}\label{eq:ClassicOptimization_constraints}
\mathrm{s.t.} \quad \tau_d = J(\theta) T + s, \\
\end{equation}
where the cost function consists on :
\begin{enumerate}
    \item $T^{T}WT$ : a quadratic term representing the power consumption of the n actuators;
    \item $s^T Q s$ : a quadratic term that allows to tune the magnitude of the slack $s$. \textcolor{black}{The slack variable allows to ensure feasibility of the optimization problem for all $\tau_d$ and define how close to $\tau_d$ the produced torques/forces will be. A big $Q$ factor is usually selected to make this slack small enough such that the solution is close enough to the controller demands while ensuring a solution at each time instant; }
    \item $\frac{\varrho}{\epsilon +\det(J(\theta)J^T(\theta))}$ : a nonlinear term which penalizes coming close to the singular configurations which reduces the rank of $J(\theta)$;
\end{enumerate}
Such a problem can be enhanced with an extra quadratic term for the angular variation and various constraints depending on the type and performances of the actuators such as thrust rate, thrust saturations, angle rate and forbidden constraints.
A major problem of this constrained optimization problem is its non-convexity due to the equality constraint in \eqref{eq:ClassicOptimization_constraints} and the penalty in \eqref{eq:ClassicOptimization_cost}. The penalty $\frac{\varrho}{\epsilon +\det(J(\theta)J^T(\theta))}$ was introduced in \cite{johansen2004constrained}, based on the fact that $\det(J(\theta)J^T(\theta))$ comes close to 0 near the rank-deficient singularities. This method is still largely used in the marine literature (\cite{martinsen2019autonomous}). The main issues with this approach is that (i) it does not work for all configurations with less vectored thrusters than DOFs ($\forall n > m : \det(J(\theta)J^T(\theta)) = 0$) and (ii) it does not provide a systematic solution everywhere, especially around $\tau_d = 0$. While nonlinear problems can be solved by using sequential programming by the means of local approximations at each time instant (\cite{johansen2004constrained}-\cite{johansen_control_2013}) and/or by considering small variations of the thrusts and angles (\cite{scibilia_constrained_2012}), issues (i)-(ii) are usually solved by ad-hoc methods. This section describes how the closed-form continuous solution presented at the previous section can be exploited also to build a smoothing cost function to be used in optimization-based CAP solutions.

The solution proposed in this section is based on the optimization formulation proposed in   \cite{ruth2007thrust} where the problem \eqref{eq:ClassicOptimization_cost}-\eqref{eq:ClassicOptimization_constraints} is rewritten using the formalism \eqref{eq:extendedMap} instead of \eqref{eq:Jtheta}. The main idea of our approach is to add the term $q_1 b(\tau_d)\|K_b^TF-q_2\|^2_2$ in the cost function:  when needed, this induces a null-space action in the directions contained in $K_b$, which is enough to stay away from singular configurations as we proved in Proposition 1 and 2. \textcolor{black}{$q_1$ and $q_2$ are scaling factors : $q_1$ is a weighting parameter within the cost function working similarly to $\rho$ in the nonlinear penalty of \eqref{eq:ClassicOptimization_cost} : the bigger, the further will the solution be from singularity configurations and thus the bigger the vehicle's maneuverability. This comes at the expense of higher power consumption. On the other hand, $q_2$ is a tuning scalar of the minimum thrust in the rest directions can be. Therefore, there is a trade-off between maneuverability (high $q_1,q_2$) and optimal power consumption (low $q_1,q_2$). }The resulting optimization problem is : 
\begin{align}
\label{eq:CVXOPT_cost}
& \min_{F,s} \quad F^{T}WF + s^TQs + q_1 b(\tau_d)\|K_b^T F - q_2\|_2^2 \\
&\text{s.t} \quad \tau_d = MF + s,
\label{eq:CVXOPT_constraints}
\end{align}

where $F^{T}WF$ represents the energy of the actuation and $s^{T}Qs$ is a slack penalty as defined in \eqref{eq:ClassicOptimization_cost}. 

The main advantage of this formulation is that it results in a convex optimization problem. Various types of constraints can be added to this formulation preserving convexity included:
\begin{itemize}
\item Saturation constraints on thrust, $\|F_i\| \leq T_{i,max};$
\item Rate constraints on thrust reusing the previous solution $\|F_i\|_0$ , $|\,\|F_i\|-\|F_i\|_0\,| \leq \Delta T_{i,max}$;
\item Pointing constraints and angular rate constraints by defining suitable $m$ convex cones $C_i$ and adding constraints in the form $F_i \in C_i$. \textcolor{black}{The cone $C_i$ can be described as an equality constraint $F_i = \sum_{j=1}^l \lambda_{ij}v_j$, i.e the result of a conic combination of $l$ vectors defining a cone, or inequality constraints using half-space formulation $AF \leq b$ as in \cite{ruth2007thrust}}.

\end{itemize}
\textcolor{black}{Depending on the constraints considered, the resulting problem is a quadratically constrained quadratic problem (QCQP) at the most. Note that, in case one needs to solve \eqref{eq:CVXOPT_cost}-\eqref{eq:CVXOPT_constraints} at high rates, it is always possible to approximate the quadratic constraints to end up with a QP that is much more easier to solve in real-time. For example, at the cost of extra conservativeness, one could rewrite a thrust saturation constraint as 3 box constraints for each $i^{th}$ actuator :
\begin{align}
    F_{x_{min}} \leq F_{x_i} &\leq F_{x_{max}}\\ 
    F_{y_{min}} \leq F_{y_i} &\leq F_{y_{max}}\\
    F_{z_{min}} \leq F_{z_i} &\leq F_{z_{max}} 
\end{align}}
\section{Numerical examples}\label{sec:numEx}
\subsection{Application of the Lipschitz continuous control allocation}
On these 2 following sections, we validate and assess the systematic behaviour of our closed-form solution \eqref{eq:OurSolutionLCAP} around singular points and its ability to generate Lipschitz continuous orientations references for the actuators. As \eqref{eq:OurSolutionLCAP} is a solution to the unconstrained CAP, we neglect the actuator constraints. These are considered in section 5.2 
\subsubsection{A 3DOF \textcolor{black}{unmanned surface vessel (USV)} with 3 azimuth thrusters}\label{sec:LipschitzUSV}
\begin{figure}[h!]
    \centering 
    \includegraphics[width = 0.7\columnwidth]{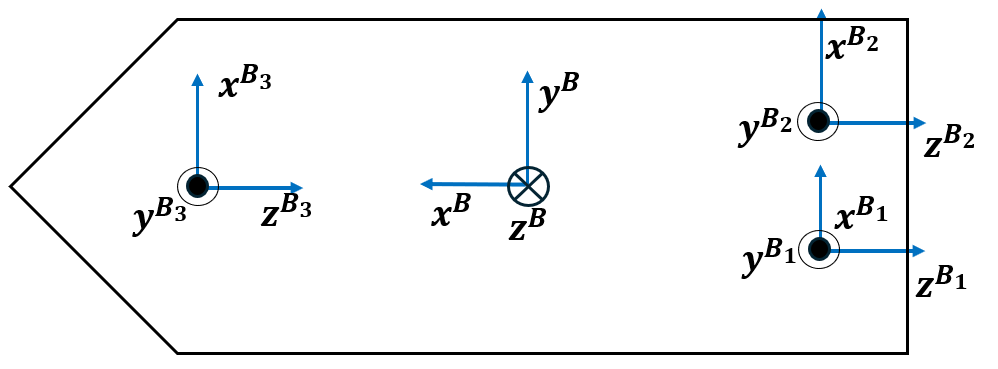}
    \caption{Surface vessel with 3 azimuth thrusters - Body fixed frame \textbf{$\boldsymbol{B}$} and the actuator-located \textbf{$\boldsymbol{B_i}$} frames.}
    \label{fig:USV:3AzSV}
\end{figure}

\begin{figure}[h!]
    \centering 
    \includegraphics[width = 0.5\columnwidth]{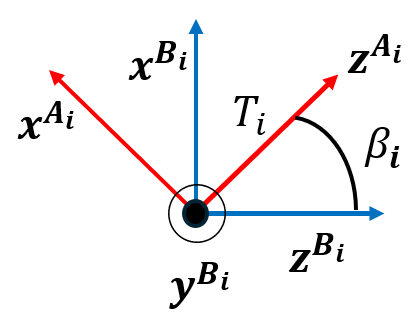}
    \caption{Surface vessel with 3 azimuth thrusters - The actuator-fixed frames \textbf{$\boldsymbol{A_{i}}$} associated to the azimuth angles $\beta_i$ produced by a rotation along the actuator-located $y^{\boldsymbol{B_i}}$ axis. The thrust is produced along the actuator-fixed axis $z^{\boldsymbol{A_{i}}}$. The rotation produced a force in the $x$ and $y$ body-framed axes.}
    \label{fig:USV:3AzSVFrames}
\end{figure}

\begin{figure}[h!]
    \includegraphics[scale=1.5,width = \columnwidth]{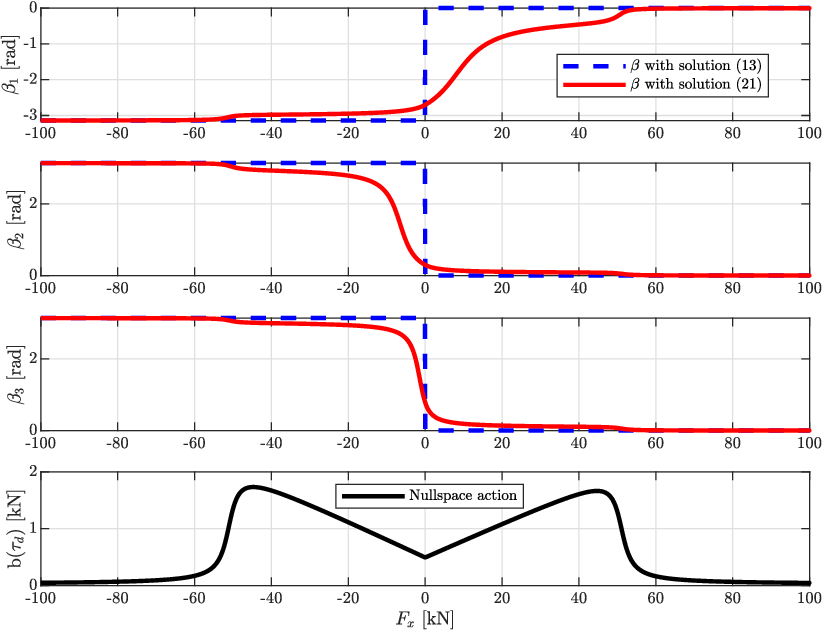}
    \caption{Evolution of the azimuth angles $\beta_i$ in the forces-torques space $\mathcal{T}_d$. Dashed curves are azimuth angles references with solution \eqref{eq:pseudoInv}, solid curves are  with solution \eqref{eq:OurSolutionLCAP}. The solid black curve is the value of the thrust \eqref{eq:Btaud} added in the nullspace direction to perform the smooth allocation}
    \label{fig:USV:LCAP}
\end{figure}

We consider a supply vessel moving in 2D with a length of 76.2 $\mathrm{m}$ and a beam of 19.6 $\mathrm{m}$ from \cite{fossen_mss_2023}. Its body frame $\boldsymbol{B}$ is the one attached to the coordinate origin of the vessel. The vessel is equipped with $m = 3$ azimuth thrusters located at positions $p_{\boldsymbol{A_{i}}}^{\boldsymbol{B}}$ = $[-30,-8,5]^T$, $[-30,8,5]^T$ and $[30,0,5]^T \mathrm{m}$ as depicted in figure \ref{fig:USV:3AzSV}. Considering a fixed elevation angle of $\frac{\pi}{2} rad$, varying the azimuth angle allows to project the thrust $T_i$ along the $x^{\boldsymbol{B}}$ and $y^{\boldsymbol{B}}$ axes into a surge $F_{x_i} = T_i cos(\beta_i)$ and sway force $F_{y_i}= T_i sin(\beta_i)$ as can be seen on Figure \ref{fig:USV:3AzSVFrames}. One can rewrite the force vector as $F = [F_{x_i}F_{y_i}]^T$ and fall back in the formalism of equation \eqref{eq:extendedMap}.
\\
For this example, let us consider an autonomous operation. The surface vessel is controlled by a PD controller designed following nonlinear PID guidelines \cite{FossenBook} (Appendix \ref{AppendixB}). The initial state is $[0\mathrm{~m},0\mathrm{~m},0\mathrm{~rad}]$ the reference is $[20\mathrm{~m},0\mathrm{~m},0 \mathrm{~rad}]$. By following equations \eqref{eq:MIMatrix}-\eqref{eq:TructatingP}, with $P = [e_{1}\, e_{2} \,e_{6}]$  and neglecting $\kappa_{d_i}$, the mapping matrix $M \in \mathbb{R}^{3\times6} $  of equation \eqref{eq:extendedMap} can be written as :
\begin{equation}\label{mappingM}
M=\left[ 
\begin{array}{cccccc}
1 & 0 & 1 & 0 & 1 & 0 \\
0 & 1 & 0 & 1 & 0 & 1 \\
-l_{y_1} & l_{x_1} & -l_{y_2} & l_{x_2} & -l_{y_3} & l_{x_3} \\
\end{array}
\right]
\end{equation}\\
where $l_{y_i}$ and $l_{x_i}$ are the $x-y$ coordinates of the position vector $p_{A_i}^B$ of the $i^{th}$ actuator. Being a constant matrix, one can compute offline once and for all its pseudo-inverse $M^\dagger \in \mathbb{R}^{6\times3}$, its associated nullspace and from it, the best smoothing directions $K_b \in \mathbb{R}^{6\times1}$ by solving \eqref{eq:BestDirection}. For the current example depicted at Figure \ref{fig:USV:3AzSV}, $K_b$ = $[0.8517,0.5241,-0.9731,$ $-0.2290,0.7071,-0.7071]^T$. With the smoothing function $b(\tau_d)$ defined in \eqref{eq:SmoothingFunction} with tuning parameters $k_a = 0.05, k_b = 0.005$ and $\varepsilon^*_{2} = 5e3$, we obtain the results of Figure \ref{fig:USV:LCAP}.

\begin{figure}[h!]
    \includegraphics[width = \columnwidth,height = \columnwidth]{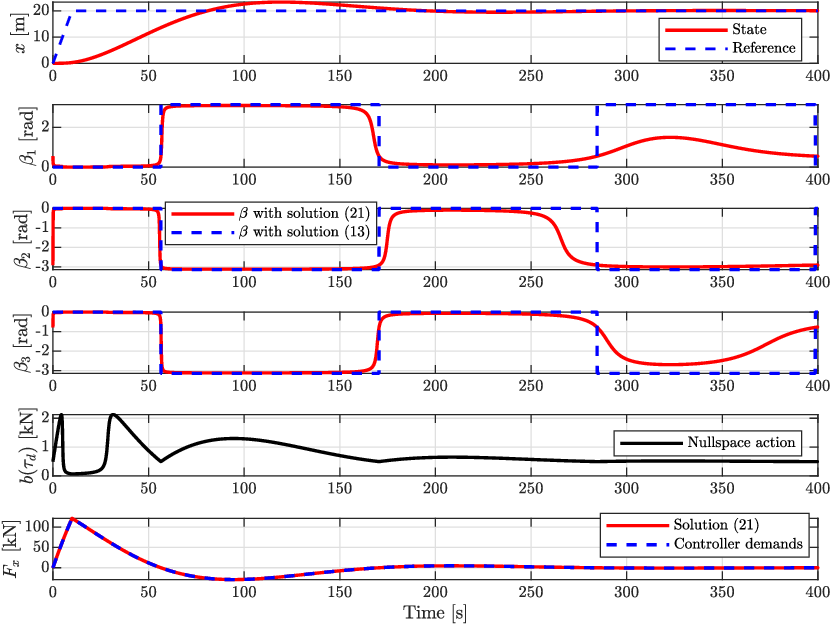}
    \caption{Position along the $x$ axis, azimuth angles $\beta$, nullspace action and surge force demand in function of time. Top plot is the $x$ position, the 3 following are the azimuth angles, the fourth one is the magnitude of the nullspace action \eqref{eq:Btaud} and the bottom plot is the desired surge force.}
    \label{fig:USV:Timeplot}
\end{figure}

As observed in Figure \ref{fig:USV:Timeplot}, the smoothing action appears when a discontinuity in azimuth orientation is close. Where solution \eqref{eq:pseudoInv} presents a jump from $\pm{\pi}$rad to 0, with the action along the nullspace dictated by eq. \eqref{eq:Fker} and \eqref{eq:SmoothingFunction}, each actuator anticipate the change of orientation. Solution \eqref{eq:OurSolutionLCAP} allows to benefit from the pseudo-inverse method away from the singular point as it can be seen on Figure \ref{fig:USV:LCAP}. The nullspace action along $K_b$ forces the actuator to rotate when coming close to the singular point. The nullspace action can be made gradually more/less present by increasing/decreasing the smoothing threshold $\varepsilon^*_2$. By \eqref{eq:btauMechanism}, the bigger $\varepsilon^*_2$, the sooner the nullspace action occurs, the less optimal the solution $F$ and vice-versa. To conclude, using \eqref{eq:upper-bound} and computing beforehand the maximum $ \varepsilon_1$ over the compact set $\mathcal{T}_d$, one can compute the maximum upper-bound to the $m$ Lipschitz constants $\max(L) =0.95 \frac{rad}{kN}$.

\subsubsection{A 6DOF quadcopter \textcolor{black}{unmanned aerial vehicle (UAV)} with tilting rotors}\label{sec:LipschitzUAV}
\begin{figure}[h!]
    \centering 
    \includegraphics[scale = 0.35]{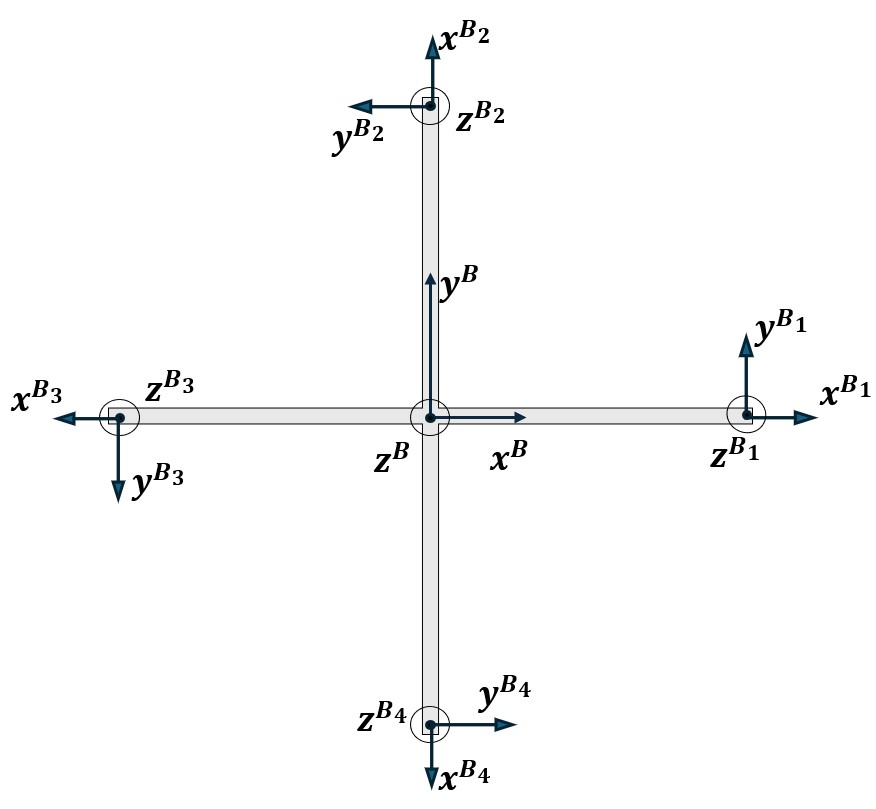}
    \caption{Representation of a X-shaped quadcopter with tilting rotors}
    \label{fig:QuadCopter}
\end{figure}
\begin{figure}[h!]
    \centering
    \includegraphics[width = 0.7\columnwidth]{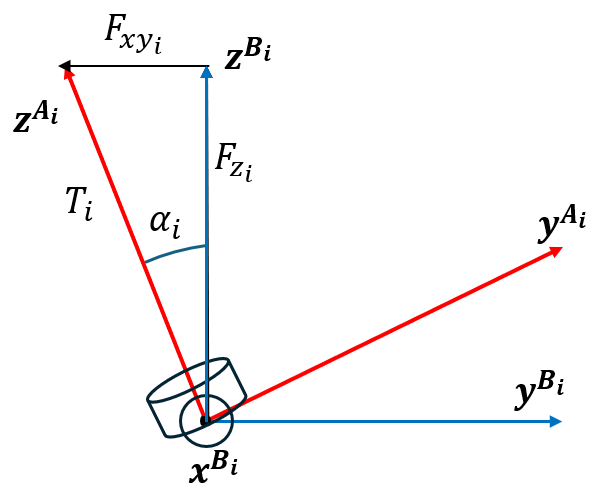}
    \caption{Representation of the actuator-fixed frame $\boldsymbol{A_i}$ and the actuator-located frame $\boldsymbol{B_i}$ -  Projection of the thrust $T_i$ on the $xy$ plane and $z$ direction by a variation of the elevation angle $\alpha_i$ to generate the projected forces $F_{xy_{i}}$ and $F_{z_i}$. Following Figure \ref{fig:QuadCopter}, the tilting introduces either a force in $x$ or $y$ axis}
    \label{fig:projFig}
\end{figure}

 To consider the extension of our methodology to an other type of rigid-body, an X-shaped quadcopter (Figure \ref{fig:QuadCopter}) with tilting propellers-rotors units actuated by servomotors (Figure \ref{fig:projFig}) is considered. These allows to produced an thrust $T_i$ in a direction $\theta_{i} = [\alpha_{i}, \beta_{i}]$ for each of the $i^{th}$ unit. The body frame $\boldsymbol{B}$ is attached at the center of the X arms, the frame $\boldsymbol{B_i}$ is located at $p_{\boldsymbol{A_{i}}}^{\boldsymbol{B}}$ with an fixed azimuth angle $\beta_i$ and the frame $\boldsymbol{A_i}$ is an actuator-fixed frame oriented by a varying elevation angle $\alpha_i$ as represented on Figure \ref{fig:QuadCopter} and \ref{fig:projFig}. Each arm is $l = 0.2$$\mathrm{m}$ long, therefore the actuators are located at $[l,0,0.05]^T$, $[0,l,0.05]^T$,$[-l,0,0.05]^T$ and $[0,-l,0.05]^T$$\mathrm{m}$ with an fixed azimuth vector $\beta = [\frac{\pi}{2},{0},{-\frac{\pi}{2}},{\pi}]^T$$\mathrm{rad}$. The overall mass of the UAV is 1 kg.
 
Following the solution \eqref{eq:OurSolutionLCAP}, one can come up with a closed-form solution to the allocation problem by following Remark 1 and adding linear equality constraints  $A_iF_i = 0$ to the actuators, enforcing a fixed azimuth $\beta_i$. One can select $A_i = [1,0,0]^T$ for actuator 1 and 3 to produce only thrust in $y^{\boldsymbol{B}}$ and $A_i = [0,1,0]^T$ 2 and 4 only in $x^{\boldsymbol{B}}$. By solving \eqref{eq:BestDirection}, we find $K_b$ = $[0, 1, 0,-1,0, 0,0, -1, 0,1,0, 0]^T$.
The simulation scenario considered here to showcase solution \eqref{eq:OurSolutionLCAP} is a rapid change of pitch torque requested by a human operator using a RC controller. The same smoothing function $b(\tau_d)$ as for the USV example is used with the following parameters : $k_a$ = 0.1, $k_b$ = 1, $\varepsilon^*_2$ = 3. To observe the effect of our solution, let us analyse the evolution of the elevation angles in function the requested pitch torque $\tau_y$ from -4Nm to 4Nm.

\begin{figure}[h!]
    \centering
    \includegraphics[width = 1\columnwidth]{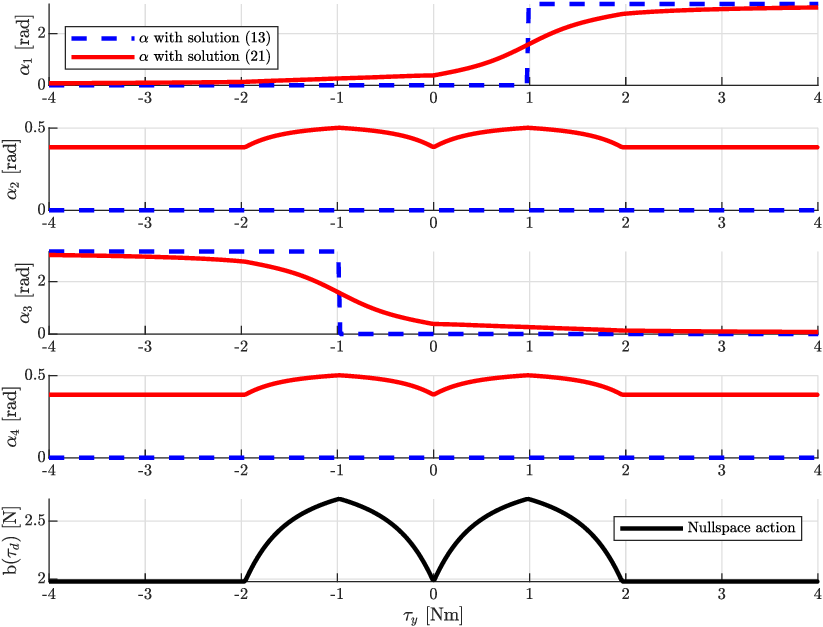}
    \caption{Evolution of the elevation angle references $\alpha_i$ in function of the demanded pitch torque. Dashed blue curves are angles with solution \eqref{eq:OurSolutionLCAP}, solid red curves are with solution \eqref{eq:OurSolutionLCAP}. The solid black curve is the value of the thrust \eqref{eq:Btaud} added in the nullspace direction to perform the smooth allocation described in Proposition 2}  
    \label{fig:UAV:QDSmooth}
\end{figure}
As illustrated in Figure \ref{fig:UAV:QDSmooth}, acting along the nullspace induces continuous variation in all tilting angles over time. We can observe a constant action due to constant gravity compensation. Moreover, we can see that nullspace thrust presents 2 maxima, which occurs whenever one elevation reference presents a discontinuity. The closer to the discontinuity, the bigger the smoothing action becomes. With the weight $w$ of the UAV being equal to $9.81 N$, these discontinuities occur whenever the pilot or the controller would request a pitch torque above $\frac{wl}{2}$, which would require to not anymore only produce a force difference to produce the torque but to exploit the force opposition. This action along the nullspace is the key to provide Lipschitz continuity of allocated angles at all times and therefore preserving overactuation. 
By computing the biggest upper bound of the $m$ Lipschitz constants $L$ for the $m$ actuator orientations, we obtain a value of $2.6834 \frac{rad}{Nm}$.

\subsubsection{Discussion}
To analyse the tradeoff between Lipschitz continuity and power consumption, we introduce the total integrated power ratio (\textit{TIPR}).  Given a set of parameters $\lambda$ of the smoothing function, the TIPR over a subdomain $\mathcal{T}_r \subset \mathcal{T}_d$ of size $r$ is defined as : 
\begin{equation}
TIPR(\lambda) = \frac{\int_{\mathcal{T}_r} \|T_{LCA}(\tau_d, \lambda)\|^2_2 d\tau_d}{\int_{\mathcal{T}_r} \|T_{PINV}(\tau_d)\|^2_2 d\tau_d}
\end{equation}
with $T_{LCA}$ being the allocated thrust with our solution \eqref{eq:OurSolutionLCAP} and $T_{PINV}$ with the pseudo inverse \eqref{eq:pseudoInv}. we approximated the integral by sampling $N=5000$ evenly spaced points of $\mathcal{T}_d$ over a subdomain of size $r=100$ and computed the allocated angles and thrusts for a different set of parameters $\lambda$ increasing the nullspace action. 
\begin{figure}[h!]
    \centering
    \includegraphics[width=\columnwidth]{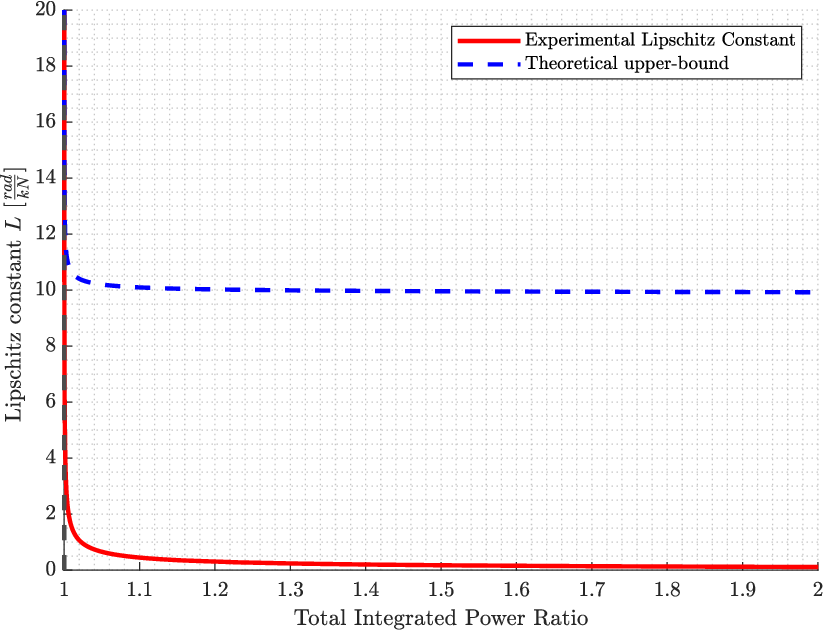}
    \caption{Maximum experiment Lipschitz constant of the allocated angles in function of the Total Integrated Power Ration (TIPR) over a section of the forces and torques space $\mathcal{T}_d$ - The experimental Lipschitz constant is in solid red while its theoretical upper-bound \eqref{eq:upper-bound} is in dashed blue}
    \label{fig:LipschizVsEnergy}
\end{figure}
One can observe that the experimental rate of change is always below the theoretical upperbound \eqref{eq:upper-bound} and that the lower the Lipschitz constant, the bigger the power consumption. Vice-versa, to be close to optimality with an unity TIPR implies an infinite Lipschitz constant. Moreover, as the final result proof \eqref{eq:upper-bound} implies it, the evolution of the Lipschitz constants follows a $\frac{2}{x}$ law. The slope and scale of both curves can be tuned by designing any function $b(\tau_d)$ that fulfills Condition \textbf{C1} and \textbf{C2}.
\subsection{Application of the dynamic optimization for constrained control allocation}
In the following section, results of the optimization problem \eqref{eq:CVXOPT_cost}-\eqref{eq:CVXOPT_constraints} applied to the USV and UAV of the previous sections are presented. We prove that by defining rest positions $K_{b}$ and enforcing them inside a optimization problem is a suitable solution to this constrained CAP. Since by definition, the least expensive solution in terms of power consumption is the one of the pseudo-inverse, solution \eqref{eq:CVXOPT_cost}-\eqref{eq:CVXOPT_constraints} is compared to the solution \eqref{eq:pseudoInv}. Furthermore, the results are presented alongside a solution from the literature (equation \eqref{eq:ClassicOptimization_cost}-\eqref{eq:ClassicOptimization_constraints} from \cite{johansen2004constrained}) for the USV case. For the UAV case, our solution \eqref{eq:CVXOPT_cost}-\eqref{eq:CVXOPT_constraints} is compared to the the pseudo-inverse \eqref{eq:pseudoInv} in a closed-loop context. The optimization problem consider as constraints the thrust saturation of the actuators $T_{max}$ and the angle rate $\Delta\alpha$ or $\Delta\beta$ to constraint the produced thrust $T_i$ in a cone $C_i$ at each sample time.  
\subsubsection{Implementation details}
\textcolor{black}{Regarding implementation, the above problems \eqref{eq:CVXOPT_cost}-\eqref{eq:CVXOPT_constraints}, including problem \eqref{eq:BestDirection}, were solved using the \textit{MPT3} toolbox with \textit{MATLAB R2024b} running on an Intel Processor 11th Gen Intel(R) Core(TM) i7-1165G7 with 16GB of RAM. Problem (32) was solved using the \textit{bnb} solver for a 1.1 seconds average while the optimization-based control allocation was solved using the solver \textit{SeDuMi-1.3} with a 10 ms average for the USV and a 13.2 ms average for the UAV. As it is run on \textit{MATLAB}, these values can be considered as worst-case scenario. Note that for the USV, as the dynamics are slow and the control loops run usually up to $10\mathrm{Hz}$, these runtimes are not critical. For the UAV case, QP approximations can be implemented and one can look up the vast and extensive literature of QCQP/QPs solved with high performance solvers (such as \textit{Gurobi}, \textit{HPIPM} or \textit{OSQP}) that run real-time \cite{chen_real-time_2020} \cite{osqp}}.
\subsubsection{A 3DOF unmanned surface vessel (USV) with 3 azimuth thrusters}
In this example, the USV of section \ref{sec:LipschitzUSV} is considered. The simulation scenario here is a human operator requesting a growing surge force through the throttle. This can be modelled by a sinusoidal surge force demand. The forces and torques vector $\tau_{d}$ to be produced varies as a sine wave of frequency $f$ = 0.01 $\unit{Hz}$ and of amplitude of A = 100 $\unit{\kilo\N}$ for $t$ = 100 seconds. The interest of such a maneuver is to study the effectiveness around the singularity $\tau_d =0$ where the azimuth \eqref{eq:azimuth} is not defined.

We consider as motors parameters : the maximum and minimum azimuth rate of change per second  as $\Delta\beta_{max} = 0.4363$ rad/s, $\Delta\beta_{min} = -\Delta\beta_{max}$ and the maximum thrust per actuator as $T_{max} = 68 \unit{kN}$. The elevation $\alpha$ stays fixed at $\frac{\pi}{2}$ rad. Tuning parameters for this simulation are : $k_a$ = 0.008, $k_b$ = 0.1, $\varepsilon^*_2$ = 4500, $W$ = $2I_{6x6}$, Q = $2\times10^4 I_{6x6}$ , $q_1 = 0.1$, and $q_2=10^4$. For solution \eqref{eq:ClassicOptimization_cost}-\eqref{eq:ClassicOptimization_constraints}, tuning parameters are Q = $2\times 10^4I_{3\times3}$, W = $2I_{3\times3}$, $\Omega$ = $10^4 I_{3\times3}$, $\varrho = 3000$ and $\epsilon = 3e^{-10}$. For both cases, the initial conditions are the thrusters at rest and pointing stern. Results for solution \eqref{eq:CVXOPT_cost}-\eqref{eq:CVXOPT_constraints} are shown alongside those of \eqref{eq:ClassicOptimization_cost}-\eqref{eq:ClassicOptimization_constraints} and \eqref{eq:pseudoInv}.

\begin{figure}[h!]
    \centering
    \includegraphics[width = \columnwidth,height = \columnwidth]{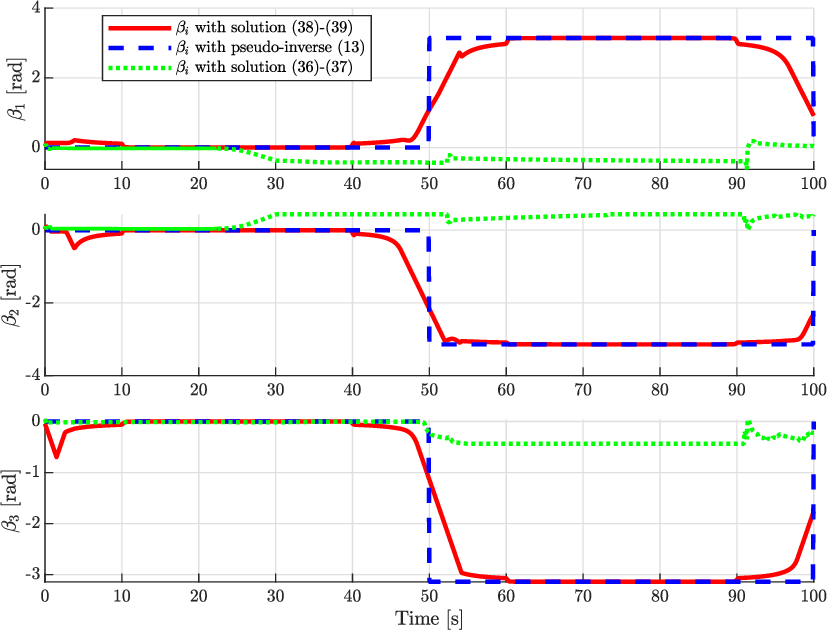}
    \caption{Evolution of the azimuth angles references $\beta_i$ over time. Green dotted curve is with solution from literature \eqref{eq:ClassicOptimization_cost}-\eqref{eq:ClassicOptimization_constraints}, blue dashed curve is with the pseudo-inverse \eqref{eq:pseudoInv} and red solid curve is with the solution \eqref{eq:CVXOPT_cost}-\eqref{eq:CVXOPT_constraints}}
    \label{fig:USV:AllTheta}
\end{figure}

\begin{figure}[h!]
    \centering
    \includegraphics[width = 1\columnwidth,height = \columnwidth]{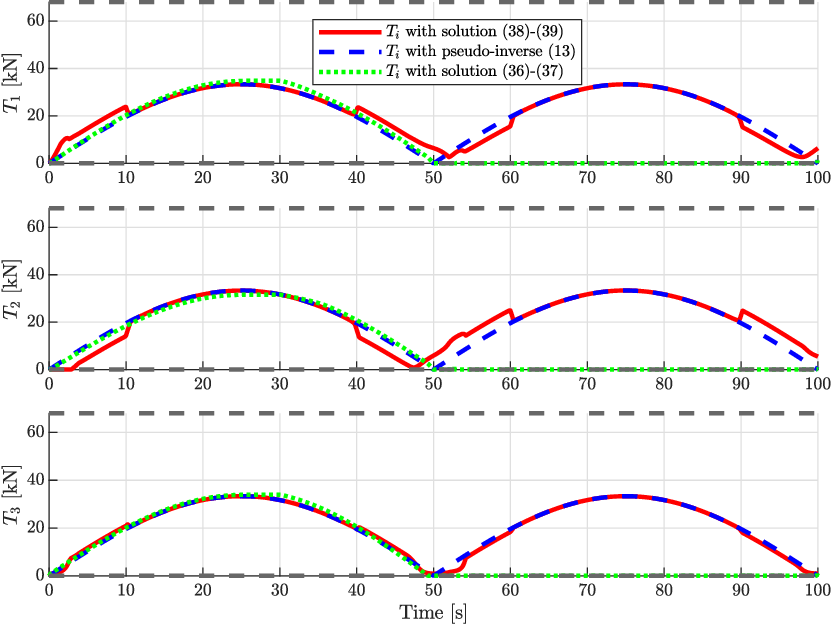}
    \caption{Evolution of the thrusts references $T_i$ over time. Green dotted curve is with solution from literature \eqref{eq:ClassicOptimization_cost}-\eqref{eq:ClassicOptimization_constraints}, blue dashed curve is with pseudo-inverse \eqref{eq:pseudoInv} and red solid curve is with the solution \eqref{eq:CVXOPT_cost}-\eqref{eq:CVXOPT_constraints}. Black dashed lines are thrust saturation limits.}
    \label{fig:USV:AllThrusts}
\end{figure}

\begin{figure}[h!]
    \centering
    \includegraphics[width=\columnwidth,height=\columnwidth]{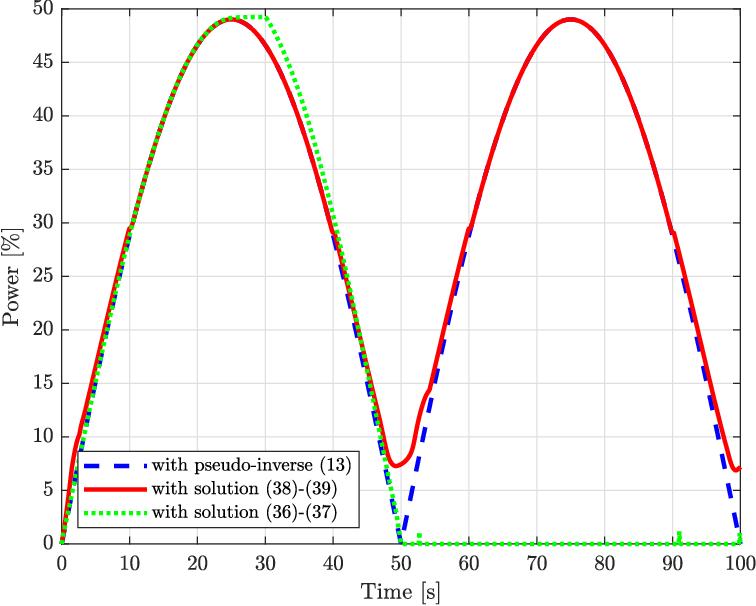}
    \caption{Evolution of the total power consumption of thrusters in \% of max power achievable over time. Green dotted curve is with solution from literature \eqref{eq:ClassicOptimization_cost}-\eqref{eq:ClassicOptimization_constraints}, blue dashed curve is with the pseudo-inverse \eqref{eq:pseudoInv} and red solid curve is with the solution \eqref{eq:CVXOPT_cost}-\eqref{eq:CVXOPT_constraints}}
    \label{fig:USV:Power}
\end{figure}

One can see on Figure \ref{fig:USV:AllTheta} that close to $\tau_d = 0$, the action along the nullspace forces the thrusters to rotate early on to reach $\pm \pi \si{rad}$ thanks to the penalty term \textcolor{black}{$q_1 b(\tau_d)\|K_b^TF-q_2\|^2_2$. The nullspace action} along $K_b$ allows the thrusters be ready to fulfill the demand of the controller after $t = 50 s$ where solution \eqref{eq:ClassicOptimization_cost}-\eqref{eq:ClassicOptimization_constraints} gets stuck in a local minima and stops producing thrusts as one can see on Figure \ref{fig:USV:AllThrusts}. This deadzone-like effect is dependant on how fast the azimuth reference varies and of course, how fast the servos are. The great advantage of solution \eqref{eq:CVXOPT_cost}-\eqref{eq:CVXOPT_constraints} is that it stays as close as possible to the pseudo-inverse one while avoiding to fall in the deadzone caused by the discontinuity of the angle reference. Indeed, one can observe on Figures \ref{fig:USV:AllThrusts}-\ref{fig:USV:Power} that between $[0,10] s$ and $[40,60] s$, the nullspace action is more and more predominant which causes a higher thrust and therefore translates into a bigger power consumption of a few \%. Away from those time intervals, the solution \eqref{eq:CVXOPT_cost}-\eqref{eq:CVXOPT_constraints} performs better than solution \eqref{eq:ClassicOptimization_cost}-\eqref{eq:ClassicOptimization_constraints}. 
\begin{figure}[h!]
    \centering
    \includegraphics[width = \columnwidth]{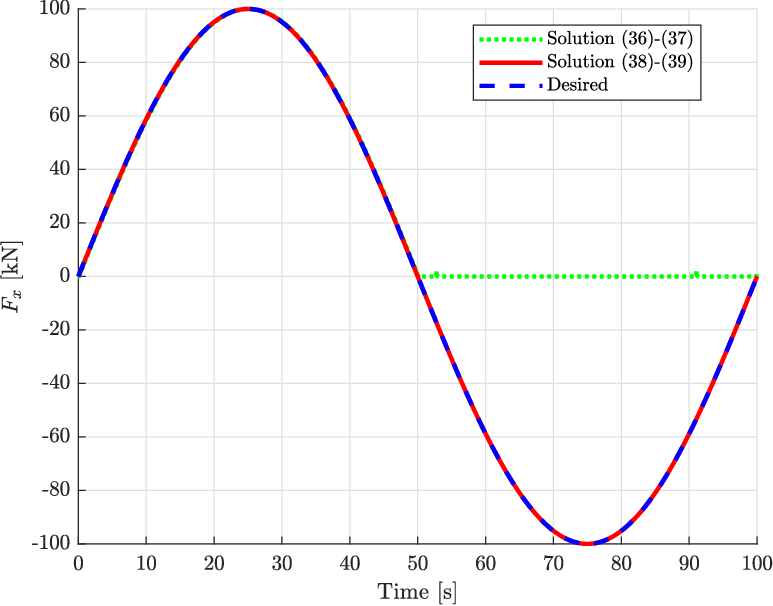}
    \caption{Evolution of $F_x$ over time - Results for a varying surge force as reference. Dashed blue line is the desired $F_x$, green dotted curve is $F_x$ with solution from literature \eqref{eq:ClassicOptimization_cost}-\eqref{eq:ClassicOptimization_constraints} and red solid curve is the produced $F_x$ with the solution \eqref{eq:CVXOPT_cost}-\eqref{eq:CVXOPT_constraints}}
    \label{fig:USV:FX}
\end{figure}

\subsubsection{\textcolor{black}{A 6DOF quadcopter UAV with tilting rotors}}
{In this case study, the quadcopter model from Section \ref{sec:LipschitzUAV} is considered with \textcolor{black}{varying azimuth and elevation vectored thrusters} in a closed-loop with cascade PD controllers for the position and attitude controller with respective gains tuned to ensure asymptotic stability of the overall closed loop system (\textcolor{black}{Appendix \ref{AppendixA}})}. \textcolor{black}{The simulation scenario is a fully autonomous operation where the UAV is brought near a kinematic singularity. By \eqref{eq:OrientationVector}, $\beta_i$ are undefined when the $\alpha_i = 0$ which can lead to hazardous $\beta_i$ causing instability. Therefore, the following experiment showcases how our method handles this gimbal singularity. While this singularity at $\alpha_i =0$ is caused by the chosen azimuth-elevation parametrization \eqref{eq:OrientationVector}, any parameter representation of a vector in the $3\text{D}$ Cartesian space by 2 angles presents kinematic singularities. With another parametrization, the issue would arise at a different point but still be present. However, our method systematically avoids them regardless of their locations.} 

We assume 4 propeller-rotors units that can each generate a maximum thrust of 10 N. The tilting units are assumed to have a maximum rate $\Delta\beta = \Delta\alpha =\pi$ rad/s. The parameters of the smoothing function $b(\tau_d)$ are $k_a = k_b = 0.5$ and $\varepsilon_2^* = 4$. The optimization problem is solved using weighting matrices $Q = 2000I_{6\times6}$, $W = I_{12\times12}$, $q_{1} = 1$, and $q_{2} = 5$. The initial conditions are provided by \eqref{eq:OurSolutionLCAP} for the optimization-based allocator.

\begin{figure}[h!]
    \centering
    \includegraphics[width = \columnwidth]{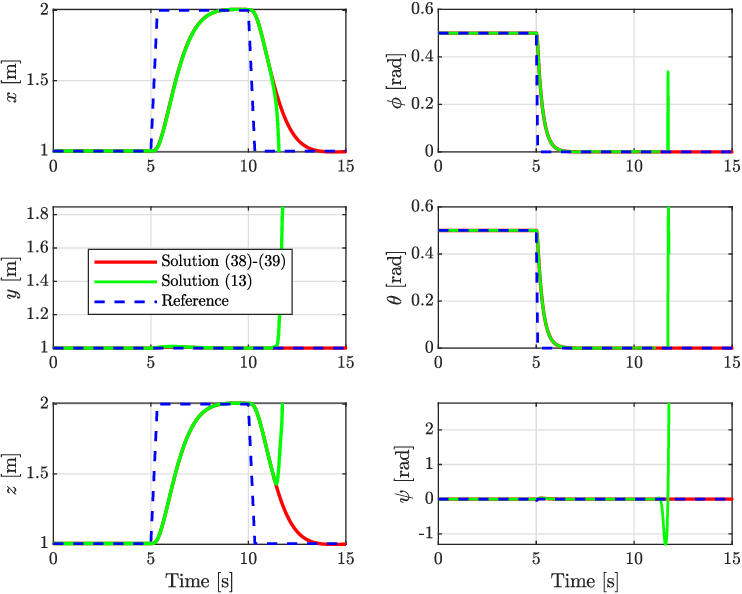}
    \caption{Evolution of states over time. Blue dashed curves are the references. 
    Red solid curves are the state evolutions with the optimization-based allocation \eqref{eq:CVXOPT_cost}-\eqref{eq:CVXOPT_constraints}. Green solid curves are the state evolutions with the pseudo-inverse based allocation \eqref{eq:pseudoInv}.}
    \label{fig:QD:AllStates}
\end{figure}

\begin{figure}[h!]
    \centering
    \includegraphics[width = \columnwidth]{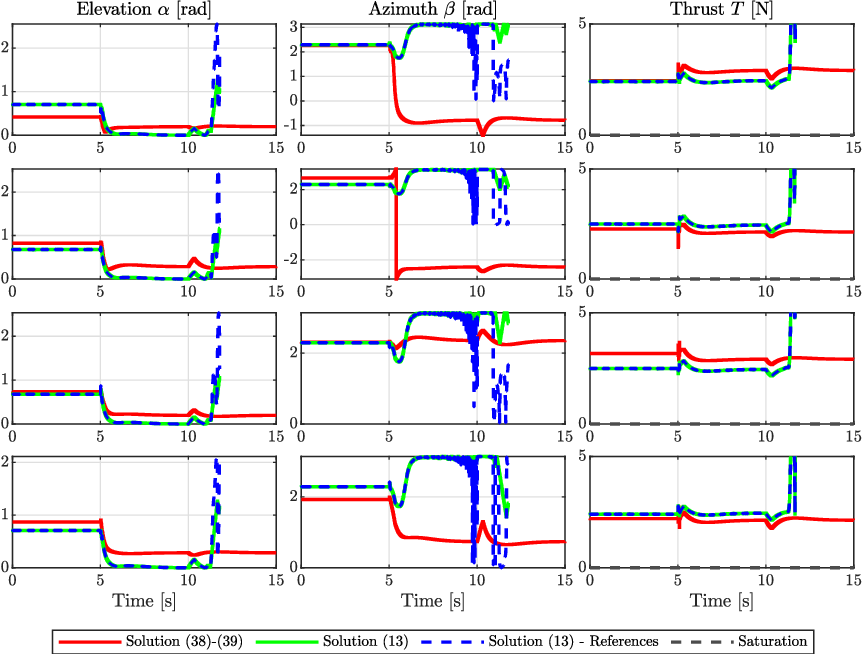}
    \caption{Evolution of the elevation angles $\alpha_i$ over time. Blue dashed curves are the allocation references from the pseudo-inverse allocation \eqref{eq:pseudoInv}, red solid curves are the resulting thrusts and angles with the optimization-based allocation \eqref{eq:CVXOPT_cost}-\eqref{eq:CVXOPT_constraints} and the green solid curves are the resulting thrusts and angles with the pseudo-inverse allocation \eqref{eq:pseudoInv}}
    \label{fig:QD:AllTheta}
\end{figure}

\begin{figure}[h!]
    \centering
        \centering
        \includegraphics[width = \columnwidth,]{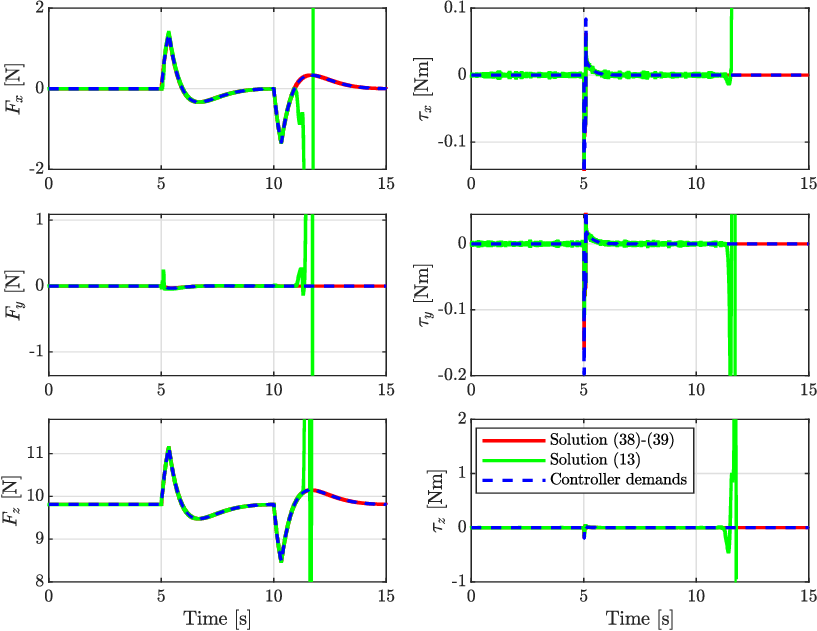}
        
    \caption{Evolution of the generalized forces $\tau$ over time. Blue dashed curves are the controller demands and red solid curves are the produced torque with the optimization-based allocation \eqref{eq:CVXOPT_cost}-\eqref{eq:CVXOPT_constraints}. Green solid curves are the forces and torques with the pseudo-inverse allocation\eqref{eq:pseudoInv}}
    \label{fig:QD:TAUS}
\end{figure}

\begin{figure}[h!]
    \centering
    \includegraphics[width = \columnwidth]{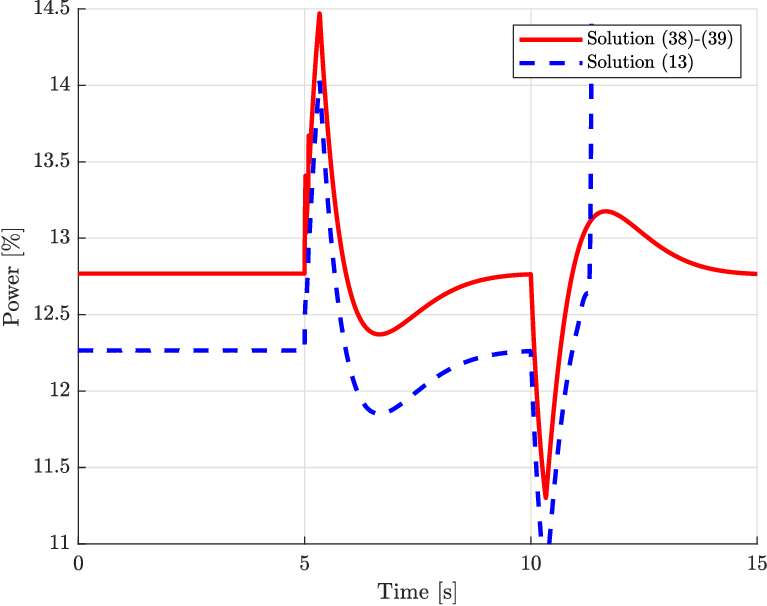}
    \caption{Evolution of the power consumption of thrusters in \% of max power achievable over time. Blue dashed curve is with pseudo-inverse allocation \eqref{eq:pseudoInv} and red solid curve is with the optimization-based allocation \eqref{eq:CVXOPT_cost}-\eqref{eq:CVXOPT_constraints}}
    \label{fig:QD:Power}
\end{figure}

Figures \ref{fig:QD:AllStates} to \ref{fig:QD:Power} show that our proposed solution, formulated in the optimization problem  \eqref{eq:CVXOPT_cost}-\eqref{eq:CVXOPT_constraints}, preserves the overactuation of the UAV by producing continuous and feasible thrust and angle references that fulfill the controller demands while staying away from singularities. In contrast, when the pseudo-inverse-based allocation (\ref{eq:pseudoInv}) is applied, instability occurs around 10 $s$ after a short period around the singularity $\alpha_i = 0$ introducing hazardous $\beta$ (see middle plot of \ref{fig:QD:AllTheta}). Note that we do not display the power and thrust curves of solution \eqref{eq:pseudoInv} after instability occurs to keep clarity of the relevant signals.

Figure \ref{fig:QD:Power} illustrates the power consumption during the maneuver. The increased power demand results from the nullspace thrust applied throughout the operation to stay away from the singularities and preserving overactuation. As explained in the previous section, the amplitude of the nullspace action can be tuned based on how low the Lipschitz constant of the allocated angles is desired.

\section{Conclusion}
In this paper, we present two key contributions to the control allocation problem (CAP) for thrust-vector-controlled rigid bodies. The first contribution is a novel closed-form solution that ensures Lipschitz continuity of thruster orientations, addressing \textcolor{black}{singular points in wrench space, rank-deficient and kinematic singularities}. The second contribution introduces a convex optimization-based formulation of the CAP, which incorporates actuator orientation rate limits and thrust saturation, while assuming negligible thrust rate due to its fast dynamics. This formulation builds on the first contribution, integrating singularity avoidance with a flexible approach that can be extended to accommodate a wide range of actuator constraints.

The effectiveness of both methods has been validated through numerical examples \textcolor{black}{showcasing the trade-off between power consumption and the magnitude of the Lipschitz constants of the actuator orientations. For the closed-form solution, a power consumption index is introduced and for the optimization-based method, a percentage of the maximum achievable power is presented. For the paper examples, up to 8$\%$ of extra power is required for ensuring continuity of the allocated angles.} \textcolor{black}{Future work will focus on extending these approaches to thrust-vector controlled multi-body systems and different scenarios involving varying actuator positions, varying smoothing direction for robust control effectiveness}.
\section*{Acknowledgements}
This research was partially supported by \textit{the MEXT SPReAD AI for Science program 2026 (Grant No. 26276994)}. The authors would like to thank Dr. Alejandro Goldar Davila for his help during the proofreading of the paper. 

\bibliographystyle{unsrt}        
\bibliography{autosam}           

\appendix
\section{Tilt-rotor quadcopter model and controller}    
We consider a nonlinear dynamic model of tilt-rotor quadcopter as follows :
\begin{equation}
\left\{\begin{aligned}
m \ddot{p}_W & =-m g \cdot e_3+R(q)F_B \\
{\left[\begin{array}{l}
\dot{q}_R \\
\dot{q}_I
\end{array}\right] } & =\frac{1}{2} E(q) \omega_B \\
J \dot{\omega}_B & =-S(\omega_B)J \omega_B+\tau_B,
\end{aligned}\right.
\end{equation}
where $F_B = [F_x,F_y,F_z]^T$ the produced forces in the body frame $\boldsymbol{B}$,$R(q)$ the body-to-world rotation matrix based on the attitude vector $q$, $\tau_B = [\tau_x,\tau_y,\tau_z]^T$ the body torques such that the generalized forces vector is $\tau_d = [F_B,\tau_B]^T$. As parameters we consider $m = 1 \mathrm{~kg}$, $g = 9.81 \mathrm{~m} / \mathrm{s}^2$ is the gravitational acceleration oriented in $e_3 = [0,0,1]^T$, $p_W \in \mathbb{R}^3$ is the UAV position in the world frame, $q = [q_R,q_I]\in \mathbb{H}$ is its attitude with $\mathbb{H}$ the set of unit quaternions, $\omega_B \in \mathbb{R}^3$ the body angular velocity and $J = \text{diag}(0.0041,0.0041,0.0082) \mathrm{~kg~m^2}$ is the inertia matrix. $E(q)$ is the differential kinematics matrix : 
\begin{equation}
E(q)=\left[\begin{array}{c}
-q_I^T \\
q_R I_{3\times3}+S(q_I)
\end{array}\right] .
\end{equation}
\label{AppendixA}
With $\tilde{q}_I$ the attitude error in body frame and $e$ the position error in the world frame, the attitude controller is $\tau=-h_p \tilde{q}_I-h_d \omega$ with $h_p = 15$,$h_d = 0.72$ and the position controller, $F_B=R(q)^T(m\left(k_pe-k_d \dot{p}+g \cdot e_3\right))$ with $k_p = 4$, $k_d = 3$ tuned to ensure asymptotic stability in the absence of constraints following  \cite{tartaglione_constrained_2024}.
\section{Supply vessel model and controller}
We consider a linear surge-sway-yaw model of a supply vessel of length $76.2 \mathrm{~m}$ from \cite{fossen_mss_2023}. The kinematic and dynamic equations are as follows : 
\begin{align}
    \dot{\eta} = R(\psi)&\nu, \label{eq:BodyVesselKinematicsAfterAssumption} \\
    M\dot{\nu} + D\nu  &= \tau_d, \label{eq:BodyVesselDynamicsAfterAssumption} 
\end{align}
with the position in NED frame $\eta = [x,y,\psi]^T$, the body-frame velocity $\nu = [u,v,r]^T$, the body-to-NED rotation matrix $R(\psi)$, $M$ the inertia and added mass matrix, $D$ the linear viscous effects matrix and $\tau_d = [F_x,F_y,\tau_z]$ the body generalized forces. With $\tilde\eta$ the position error, the control law is a nonlinear PD controller :  $\tau_d = K_pR(\psi)^T\tilde\eta - K_d\nu$, where $K_p$ and $K_d$ are $3\times 3$ positive definite matrices such that (i) global asymptotic stability is ensured in the absence of constraints and environmental disturbances; (ii) the closed-loop step response presents a natural frequency $\omega = 0.1 \mathrm{~rad/s}$ and a damping ratio $\zeta = 0.8$ following \cite{FossenBook}.
\label{AppendixB}



\end{document}